\documentclass[twocolumn,aps,showpacs,amssymb,floatfix]{revtex4}
\newcommand{\be}{\begin{equation}}
\newcommand{\ee}{\end{equation}}
\newcommand{\beq}{\begin{eqnarray}}
\newcommand{\eeq}{\end{eqnarray}}
\usepackage{bm}
\usepackage{graphicx}%
\usepackage{epsf}
\usepackage{epsfig}

\usepackage{amsmath}
\usepackage{amsfonts,amssymb}

\begin{document}

\newcommand{\twopt}[5]{\langle G_{#1}^{#2}(#3;\mathbf{#4};\Gamma_{#5})\rangle}
\newcommand{\threept}[7]{\langle G_{#1}^{#2}(#3,#4;\mathbf{#5},\mathbf{#6};\Gamma_{#7})\rangle}

\title{The axial N to $\Delta$  transition form factors from Lattice QCD}
\author{C.~Alexandrou~$^a$, Th. Leontiou~$^a$, J. W. Negele~$^b$, and A. Tsapalis~$^c$}
\affiliation{{$^a$ Department of Physics, University of Cyprus, CY-1678 Nicosia, Cyprus}\\
{$^b$ 
Center for Theoretical Physics, Laboratory for
 Nuclear Science and Department of Physics, Massachusetts Institute of
Technology, Cambridge, Massachusetts 02139, U.S.A.}\\
{$^c$ 
Institute for Accelerating Systems and Applications, University of Athens,
Athens,Greece.}}

\date{\today}%

\begin{abstract}

 We evaluate the $N$ to $\Delta$ axial transition form factors
in lattice QCD 
 with no dynamical sea quarks,  with two degenerate flavors of dynamical
Wilson quarks,  and using domain wall valence 
fermions with three flavors of staggered sea quarks.
We predict the ratio $C_5^A(q^2)/C_3^V(q^2)$ relevant 
for  parity violating asymmetry experiments and verify the off-diagonal Goldberger-Treiman relation.  
\end{abstract}
\maketitle
 
The electromagnetic structure of the nucleon including the transition
form factors for electroproduction of the $\Delta$ has been the 
subject of recent
experimental~\cite{BatesClas} and theoretical 
studies~\cite{chiral,lattice}. 
The N to $\Delta$ transition has the  advantages that
the $\Delta(1232)$ is the dominant, clearly accessible nucleon resonance,
and the isovector spin-flip transition provides selective
information on hadron structure.
The weak structure functions provide valuable input
often complimentary to
that obtained from electromagnetic probes.
In the nucleon, for example,  measurements of the
elastic parity violating asymmetry yield 
information  on strange quark contributions.
 The N to $\Delta$ transition filters 
out the isoscalar  $s\bar{s}$ contributions, 
so  the parity violating asymmetry in 
weak neutral and charge changing  
N to  $\Delta$ transitions probes isovector
structure~\cite{Nimai,Wells}  not accessible 
in the study of strange isoscalar quark currents.
Furthermore its isovector nature
could be used to detect physics beyond the standard model since 
it is sensitive
to additional heavy particles not appearing in the standard model. However,
in order
to compete with other low-energy semi-leptonic measurements, 
the N to $\Delta$ transition requires a determination
of the parity violating asymmetry to significantly  better than
 1\% precision~\cite{Nimai}. 
Hence, a lattice prediction for  
the axial form factors provides valuable input 
for ongoing experiments~\cite{Wells}.
In this work, we evaluate
the dominant contribution to  the parity violating 
asymmetry, determined by the ratio $C_5^A/C_3^V$. This
is the off-diagonal analogue of the $g_A/g_V$ ratio 
extracted from neutron
$\beta$-decay and therefore  tests low-energy consequences of chiral symmetry, 
such as the
off-diagonal Goldberger-Treiman relation. In addition, the ratio of axial 
form factors
$C_6^A/C_5^A$ provides
 a measure of axial current conservation.

Since this is the first lattice
computation of the  axial N to $\Delta$ transition form factors, 
the starting point is  an evaluation in the quenched theory
using the standard Wilson action. A quenched calculation
 allows us to use a large
 lattice in order to minimize finite volume effects and obtain accurate
results  at small momentum transfers reaching  pion mass, $m_\pi$, down to
about 410 MeV.
In order to study the role
of the pion cloud, which is expected to provide
an important ingredient in the description of the
 properties of the nucleon system,
one requires dynamical configurations with light quarks on large volumes. 
In this
work,  the light quark regime is  studied in two ways.
 First, we use  configurations with the lightest available
dynamical two flavor Wilson fermions spanning approximately the same
pion mass range as in the quenched calculation~\cite{TchiL,Carsten}.
 Second, we use a hybrid combination of domain wall valence quarks, which have chiral
symmetry on the lattice, and
MILC configurations generated with three flavors of staggered sea
quarks  using the Asqtad improved action~\cite{MILC}.
The effectiveness of this hybrid combination has recently been demonstrated in the 
successful precision calculation of the axial charge,  $g_A$~\cite{gaxial}. 
Since Wilson fermions
have discretization errors in the lattice spacing, $a$, of ${\cal O}(a)$ and break chiral symmetry whereas the hybrid action has discretization
 errors of  ${\cal O}(a^2)$ and chirally symmetric valence fermions, agreement between calculations using these two
lattice actions provides a non-trivial check of consistency of the lattice results.
The hybrid calculation is the most  computationally demanding, 
 since it requires propagators on a five-dimensional lattice.
The bare quark mass for the domain wall fermions, the size of
the fifth dimension and the renormalization
factors, $Z_A$, for the four-dimensional axial current are
 taken from Ref.~\cite{gaxial}. 
As in the case  of Wilson fermions,
 we consider three values of light quark mass
with the strange sea quark mass fixed to approximately
its physical value~\cite{MILC}. In all cases we use Wuppertal 
smeared~\cite{Wuppertal} interpolating
fields at the
source and sink. In the unquenched Wilson case, 
to minimize fluctuations~\cite{nucleonff} we use hypercubic (HYP)
 smearing~\cite{HYP} on
 the spatial links
entering in the Wuppertal smearing function at the source and sink whereas
for the hybrid case all gauge links in the fermion action are HYP smeared. 
 We list the parameters used in our computations 
in Table~\ref{table:parameters}. 
The value of the lattice spacing
 is determined from the nucleon mass 
at the physical limit for the case of Wilson fermions and for the staggered
sea quark configurations,
 we take the value determined from heavy quark
 spectroscopy~\cite{MILCa}.

The invariant N to $\Delta$ weak matrix element, expressed
in terms of four transition
form factors~\cite{alder,LS} can be written as~\cite{Wells}
\small
\beq
<\Delta(p^{\prime},s^\prime)|A^3_{\mu}|N(p,s)> &=& i\sqrt{\frac{2}{3}} 
\left(\frac{M_\Delta M_N}{E_\Delta({\bf p}^\prime) E_N({\bf p})}\right)^{1/2}
\nonumber \\ 
&\>&\hspace*{-4cm}\bar{u}^\lambda(p^\prime,s^\prime)\biggl[\left (\frac{C^A_3(q^2)}{M_N}\gamma^\nu + \frac{C^A_4(q^2)}{M^2_N}p{^{\prime \nu}}\right)  
\left(g_{\lambda\mu}g_{\rho\nu}-g_{\lambda\rho}g_{\mu\nu}\right)q^\rho \nonumber \\
&\>&\hspace*{-2cm}+C^A_5(q^2) g_{\lambda\mu} +\frac{C^A_6(q^2)}{M^2_N} q_\lambda q_\mu \biggr]u(p,s)
\label{matrix element}
\eeq
\normalsize
where $q_\mu=p{^\prime}_\mu-p_\mu$ is the momentum {transfer,
 $A^3_\mu(x)=  \bar{\psi}(x)\gamma_\mu \gamma_5 \frac{\tau^{3}}{2} \psi(x)$ 
is the isovector part of the axial current, and $\tau^3$ is 
the third Pauli matrix. 
We evaluate this matrix element on the lattice by computing
the nucleon two-point function $G^{N}(t;{\bf p};\Gamma_4)$, 
the $\Delta$ two-point function $G_{ii}^{\Delta}(t;{\bf p};\Gamma_4)$, 
and the three
point function  $\threept{\sigma}{\Delta j^\mu
N}{t_2}{t_1}{p'}{p}{}$ and forming the ratio~\cite{lattice}
\small
\begin{eqnarray} \nonumber
&\>&\hspace*{-2.3cm}R_\sigma(t_2,t_1;\mathbf{p'},\mathbf{p};\Gamma;\mu)
=\frac{\threept{\sigma}{\Delta j^\mu
N}{t_2}{t_1}{p'}{p}{}}{\twopt{ii}{\Delta}{t_2}{p'}{4}} \nonumber \\
&\>& \hspace*{-2.3cm}\left[ \frac{ \twopt{}{N}{t_2-t_1}{p}{4}
\twopt{ii}{\Delta}{t_1}{p'}{4}
\twopt{ii}{\Delta}{t_2}{p'}{4}
}{\twopt{ii}{\Delta}{t_2-t_1}{p'}{4}\twopt{}{N}{t_1}{p}{4}\twopt{}{N}{t_2}{p}{4}
}\right]^{1/2} \nonumber \\
&\overset{t_2-t_1\gg1,t_1\gg1}{\Rightarrow}&\Pi_\sigma(\mathbf{p'},\mathbf{p};\Gamma;\mu),
\label{ratio}
\end{eqnarray}
\normalsize
where the indices $i$ are summed,
$\Gamma_4=\frac{1}{2}\left(\begin{array}{cc} I &0\\0 & 0 \end{array}\right)$,
and  $\Gamma_j=\frac{1}{2}\left(\begin{array}{cc} \sigma_j &0\\0 & 0 \end{array}\right)$.}
For large time separations  between $t_1$, the time when a photon interacts with
a quark,  and $t_2$, the time when the $\Delta$ is annihilated,
 the ratio of Eq.~(\ref{ratio})
becomes time independent and yields the transition matrix element of 
Eq.~(\ref{matrix element}). 
The source-sink time separation is optimized as in Ref.~\cite{nucleonff}
so that a plateau 
is clearly identified when varying $t_1$.  
We use kinematics where the $\Delta$ is produced
at rest and  $Q^2=-q^2$ is the Euclidean momentum transfer squared.
There are various choices for the Rarita-Schwinger spinor
index,  $\sigma$, and projection matrices, $\Gamma$,  that can be used in the computation of
the three point function, each requiring a 
 sequential inversion.
We use this freedom to construct
 optimized $\Delta$ sources that maximize the number of
lattice momentum vectors contributing to a given value of $Q^2$
in analogy with  the evaluation of the
electromagnetic N to $\Delta$ transition form factors~\cite{lattice}.

\begin{table}[h]
\caption{ The number of configurations,
 the hopping parameter, $\kappa$, for Wilson fermions or the light-quark mass, $m_l$, for staggered quarks, and 
 the pion,  nucleon and $\Delta$ masses in lattice
units.} 
\label{table:parameters}
\begin{tabular}{|c|c|c|c|c|}
\hline
\multicolumn{1}{|c|}{no. confs } &
\multicolumn{1}{ c|}{$\kappa$ or $am_l$ } &
\multicolumn{1}{ c|}{$am_\pi$ } &
\multicolumn{1}{ c|}{$aM_N$ } &
\multicolumn{1}{ c|}{$aM_{\Delta}$ } 
\\
\hline
\multicolumn{5}{|c|}{Quenched $32^3\times 64$ \hspace*{0.5cm} $a^{-1}=2.14(6)$ GeV}
 \\ \hline
  200            &  0.1554 &  0.263(2) & 0.592(5)   & 0.687(7) \\
  200            &  0.1558 & 0.229(2) &  0.556(6)   & 0.666(8) \\
  200            &  0.1562 & 0.192(2) &  0.518(6)   & 0.646(9)\\
    &  $\kappa_c=$0.1571   & 0.       &  0.439(4)  & 0.598(6)\\
\hline
\multicolumn{5}{|c|}{Unquenched Wilson $24^3\times 40 $~\cite{TchiL}  \hspace*{0.5cm}
$a^{-1}=2.56(10)$ GeV} 
 \\\hline
 185                &  0.1575  & 0.270(3) & 0.580(7) & 0.645(5)\\
 157                &  0.1580  & 0.199(3) & 0.500(10) & 0.581(14) \\
\hline
\multicolumn{5}{|c|}{Unquenched Wilson $24^3\times 32 $~\cite{Carsten}  \hspace*{0.5cm}
$a^{-1}=2.56(10)$ GeV} 
\\\hline
 200                &  0.15825 & 0.150(3)~\cite{footnote}
 & 0.423(7)  & 0.533(8)  \\
                    & $\kappa_c=0.1585$& 0. & 0.366(13)& 0.486(14)\\
\hline
\multicolumn{5}{|c|}{MILC $20^3\times 64 $  \hspace*{0.5cm}
$a^{-1}=1.58$ GeV} 
 \\\hline
 150                &  0.03  & 0.373(3) & 0.886(7) & 1.057(14)\\
 150                &  0.02  & 0.306(3) & 0.800(10)&  0.992(16)\\
\hline
\multicolumn{5}{|c|}{MILC $28^3\times 64 $  \hspace*{0.5cm}
$a^{-1}=1.58$ GeV} 
\\\hline
 118                &  0.01 & 0.230(3) & 0.751(7)  & 0.988(26)  \\
\hline
\end{tabular}
\end{table} 

\begin{figure}[h]
\epsfxsize=8truecm
\epsfysize=9truecm
\mbox{\epsfbox{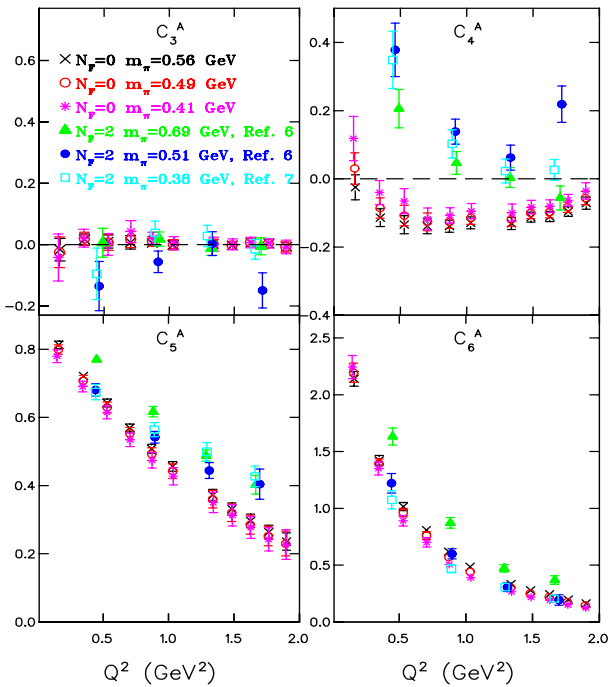}}
\caption{The axial form factors $C^A_3$, $C^A_4$, $C^A_5$ and $C^A_6$
 as a function of $Q^2$.  We show  quenched lattice results 
at $\kappa=0.1554$
 (crosses), at $\kappa=0.1558$ (open circles) and at $\kappa=0.1562$
 (asterisks) and unquenched Wilson results
 at $\kappa=0.1575$ (filled triangles),
$\kappa=1580$ (filled circles) and $\kappa=0.15825$ (open squares).
 We use $Z_A=0.8$ ~\cite{renorm,renorm_dyn}.}
\label{fig:wilson form factors}
\vspace*{-0.7cm}
\end{figure}

\begin{figure}[h]
\epsfxsize=8truecm
\epsfysize=8.5truecm
\mbox{\epsfbox{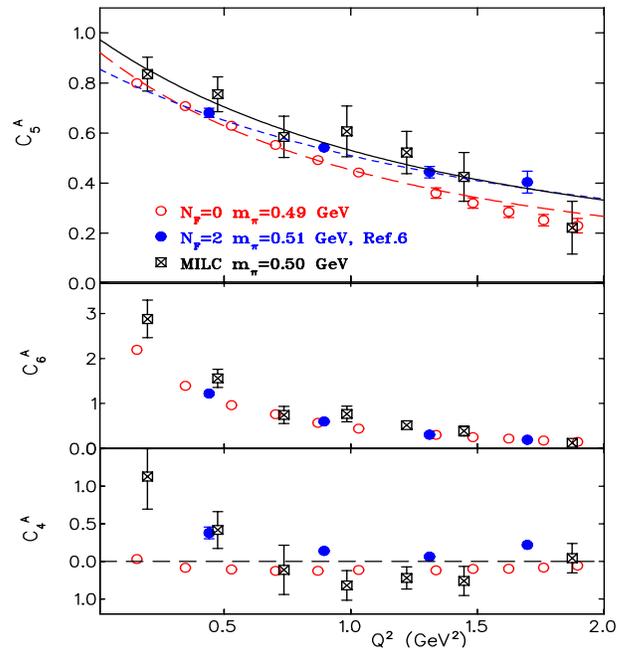}}
\caption{The dominant axial form factors  $C^A_5$ (top) and $C^A_6$ (middle)
and $C^A_4$ as a function of $Q^2$,
for similar pion mass, for quenched at $\kappa=0.1558$ (open circles ) and
dynamical Wilson fermions at $\kappa=0.1580$ (filled circles)
 and in the hybrid approach 
at $am_l=0.01$ (squares with a cross). The curves are fits to the dipole form 
$C_5^A(0)/(1+Q^2/M_A^2)^2$.}
\label{fig:CA5_6}
\vspace*{-0.2cm}
\end{figure}

In Fig.~\ref{fig:wilson form factors} we
show  quenched and unquenched results
obtained with  Wilson fermions. 
All errors shown are obtained 
using a  jackknife analysis.
 We observe that $C_3^A$ is 
consistent with zero and that unquenching effects 
are small for the dominant form factors, $C_5^A$ and $C_6^A$.
 In contrast, for the form factor $C_4^A$, dynamical fermions produce  a dramatic increase at
low momentum transfer relative to the quenched results.    Such large deviations between
quenched and full QCD results for these relatively heavy
 quark masses  are unusual,
making  this  an interesting quantity with which to study 
effects of unquenching. 
In Fig.~\ref{fig:CA5_6},
we compare the  { Wilson and hybrid results for the two dominant
form factors  for   $m_\pi \sim 500$~MeV. 
The hybrid results thus corroborate the small unquenching effects, and similar behavior is observed at the heavier and lighter quark masses.
For $C_4^A$, also shown in Fig.~\ref{fig:CA5_6}, we observe 
the same large unquenching 
effects observed for dynamical Wilsons.
In addition, although they are not shown, hybrid calculations yield $C_3^A \sim 0$. A dipole Ansatz 
$C_5^A(0)/(1+Q^2/M_A^2)^2$ 
describes  the $Q^2$-dependence of $C_5^A$ well,}
as shown by the curves in Fig.~\ref{fig:CA5_6},
yielding, at this pion mass, an axial mass
 $M_A \sim 1.8(1)$~GeV. 
In this range of
pion masses, we observed a  weak quark mass
dependence for $M_A$ that however needs to be checked
for lighter quarks before comparing to
 the experimental result of  $1.28\pm 0.10$~GeV~\cite{Kitagaki}.

\begin{figure}[h]
\epsfxsize=8truecm
\epsfysize=5truecm
\mbox{\epsfbox{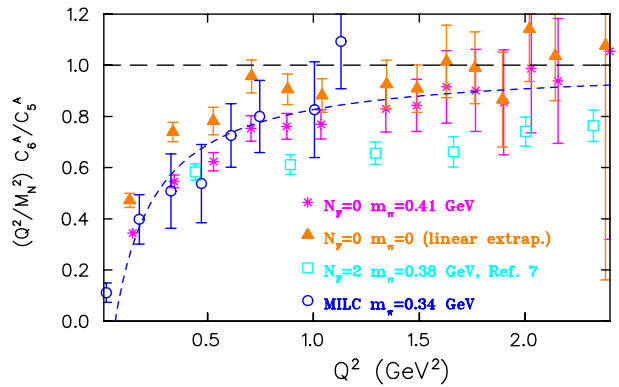}}
\caption{The ratio $\left(Q^2/M_N^2\right) C^A_6/C^A_5$ is shown versus $Q^2$
 in the quenched theory at $\kappa=0.1562$ (asterisks) and in the
physical limit (filled triangles), for 
dynamical Wilson fermions at $\kappa=0.15825$ (open squares)
 and in the hybrid approach 
at $am_l=0.01$ (open circles). {
The short dashed line denotes the soft pion fit described in the text.} }
\label{fig:light quark ratio}
\vspace*{-0.5cm}
\end{figure}

\begin{figure}[h]
\epsfxsize=8truecm
\epsfysize=5truecm
\mbox{\epsfbox{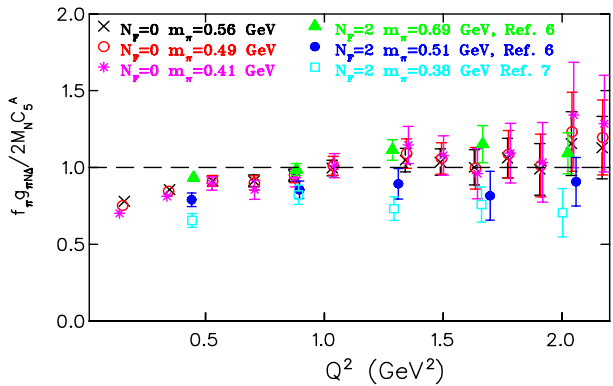}}
\caption{The ratio $f_\pi g_{\pi N\Delta}/\left(2M_N C^A_5\right)$ 
{as a function of} $Q^2$. The notation is the same as in 
Fig.~\ref{fig:wilson form factors}.
}
\label{fig:Goldberger-Treiman}
\vspace*{-0.5cm}
\end{figure}

\begin{figure}[h]
\epsfxsize=8truecm
\epsfysize=5truecm
\mbox{\epsfbox{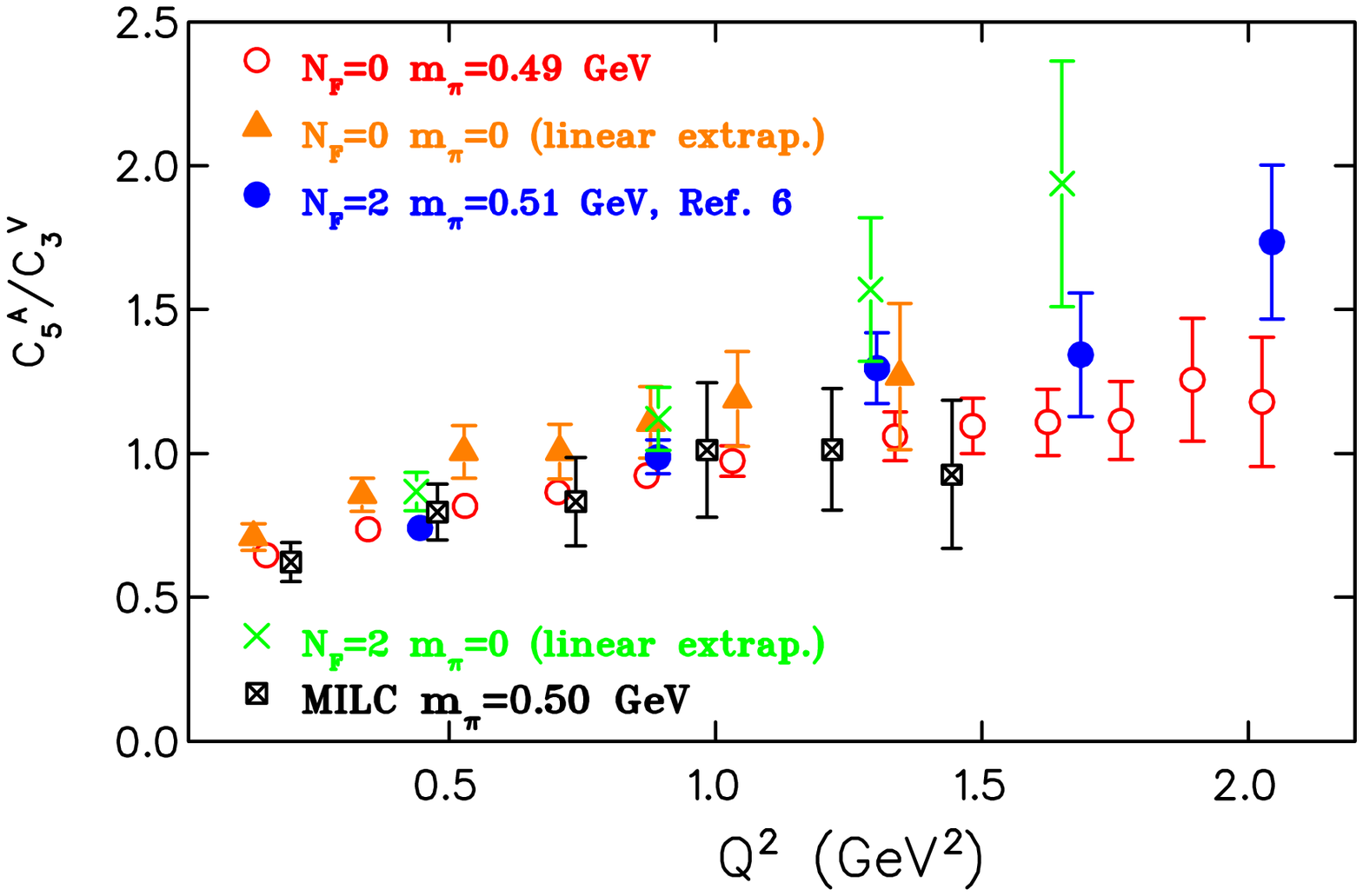}}
\caption{The ratio $C^A_5/C^V_3$ { as a function of} $Q^2$ 
for quenched QCD at $\kappa=0.1558$ (open circles) and 
at the physical {pion mass} (filled triangles), for  
dynamical Wilson fermions at $\kappa=0.1580$ (filled circles)
and at the physical  { pion mass}  (crosses)  and { for the hybrid action}
at $am_l=0.02$ (squares  with cross).}
\label{fig:CA5 over CV3}
\vspace*{-0.5cm}
\end{figure}

In the chiral limit, axial current conservation
leads to the relation
$
C_6^A(Q^2)=M_N^2 C_5^A(Q^2)/Q^2 .
$
In Fig.~\ref{fig:light quark ratio}, we show the ratio 
$\left(Q^2/M_N^2\right) C_6^A(Q^2)/C_5^A(Q^2)$ for Wilson and domain wall 
fermions
 at the lightest quark mass available
in each case. The expected value in the chiral limit for this ratio is one. 
For finite quark mass, the axial current is not conserved and for Wilson fermions
chiral symmetry is broken, so that deviations from one are expected. 
We observe  that this ratio differs from unity at low $Q^2$ but
approaches
 unity at higher values of $Q^2$. For the quenched case, where we have accurate
results, a linear extrapolation in $m_\pi^2$ yields values that
are consistent with unity for $Q^2\stackrel{>}{\sim}0.5$~GeV$^2$.
It is reassuring that this chiral restoration is seen on the lattice 
even for
Wilson fermions, confirming that the discrete theory is correctly representing 
continuum physics.
For finite pion mass,  deviations from unity
are expected to be proportional to $m_\pi^2/(Q^2+m_\pi^2)$ 
 for chiral lattice fermions. As can be
seen by the dashed line in the figure, this form 
describes well the $Q^2$-dependence of the results 
obtained with domain wall fermions.

For finite mass pions, partial conservation of {the} axial current 
($\partial_\mu A_\mu^a(x)=f_\pi m_\pi^2\pi^a(x)$) and pole dominance
lead to the
off-diagonal 
Goldberger-Treiman relation
$
C_5^A(Q^2)=f_\pi g_{\pi N\Delta}(Q^2)/2M_N
$
where $g_{\pi N\Delta}(Q^2)$ is determined
from the matrix element of the pseudoscalar density 
\hbox{$<\Delta^+|\bar{\psi}(x)\gamma_5\frac{\tau^3}{2}\psi(x)|p>$}~\cite{Liu} 
and the pion decay constant,
$f_\pi$, { is calculated} from the two-point function \hbox{$<0|A_4(x)|\pi>$}. To relate the 
lattice pion matrix element to its physical value, we need the
 pseudoscalar
renormalization constant, $Z_p$, and we use for quenched~\cite{renorm} 
and dynamical Wilson fermions~\cite{renorm_dyn} the value
\hbox{$Z_p(\mu^2a^2\sim 1)$}=0.5. 
 {In Fig.~\ref{fig:Goldberger-Treiman} we show the ratio 
$f_\pi g_{\pi N\Delta}/\left(2 M_N C_5^A\right)$ for  Wilson fermions, which is almost  independent of $Q^2$, indicating 
similar $Q^2$ dependence for $g_{\pi N\Delta}$ 
and
$C_5^A$. The ratio becomes consistent with
 unity for $Q^2\stackrel{>}{\sim} 1$~GeV$^2$, in agreement 
with the off-diagonal
 Goldberger-Treiman relation.
The large deviations seen  for the lightest quark mass for dynamical fermions
point to lattice artifacts that become more dominant for small quark masses.


{ The presently unmeasured ratio $C_5^A/C_3^V$  is an interesting prediction of our lattice calculation.}
 The form factor $C_3^V$ can be obtained from the electromagnetic 
N to $\Delta$ transition. Using our lattice
results for the dipole and electric quadrupole Sachs factors, ${\cal G}_{M1}$ and
${\cal G}_{E2}$~\cite{lat05}, we extract
$C_3^V$ using the relation 
\be
C_3^V=\frac{3}{2}\frac{M_\Delta(M_N+M_\Delta)}{(M_N+M_\Delta)^2+Q^2}
\left({\cal G}_{M1}-{\cal G}_{E2}\right). 
\ee

{  From the ratio $C_5^A/C_3^V$
shown in Fig.~\ref{fig:CA5 over CV3}, we observe that 
 quenched and unquenched results at  $m_\pi \sim 500$ MeV agree within errors,
 and that the quenched data have a sufficiently weak mass dependence 
that when extrapolated to the physical pion mass, the results nearly 
coincide with the 500 MeV data within errors. 
  Hence, it is reasonable to use the extrapolated 
quenched results as a first estimate of this ratio,   
which is the off-diagonal analogue of $g_A/g_V$ in the nucleon. 
In addition to  testing low energy consequences
of chiral symmetry~\cite{Nimai}, this ratio provides the basis for a physical estimate of the
parity violating asymmetry. Under the assumptions that
 $C_3^A\sim 0$ and 
$C_4^A$ is suppressed as
compared to $C_5^A$, both of which are supported
by our lattice results, the parity violating asymmetry can be shown to be
  proportional to 
this ratio~\cite{Nimai}. Thus, our lattice results show  that this ratio and, to 
a first approximation, the parity violating asymmetry is non-zero
at $Q^2= 0$ and increases for $Q^2\stackrel{>}{\sim}1.5$~GeV$^2$. }

In summary, we have provided a first lattice
 calculation of the axial transition form
factors, 
which are to be measured at Jefferson Lab~\cite{Wells}.
The first conclusion is that  $C_3^A$
is consistent with zero whereas $C_4^A$ is small but shows {unusually high} sensitivity to unquenching effects. The two dominant 
form factors are $C_5^A$
and $C_6^A$, { which}  are related in the chiral limit
by axial current conservation. { The
ratio $\left(Q^2/M_N^2\right) C_6^A/C_5^A$ is shown to approach unity as the
quark mass decreases, as expected from chiral symmetry.
For any quark mass, 
the strong coupling,
$g_{\pi N\Delta}$, and 
the axial form factor,
$C_5^A$,  show similar $Q^2$ dependence, and 
the off diagonal Goldberger-Treiman 
relation is reproduced as the quark mass decreases.
The ratio
of $C_5^A/C_3^V$, which provides a first  approximation to the the parity violating
asymmetry, is predicted to be non-zero at $Q^2=0$  with a two-fold increase
 when $Q^2\sim 1.5$~GeV.}

{\bf Acknowledgments:}
We thank the $T\chi L$ collaboration~\cite{TchiL}
as well as  C. Urbach {\it et al.}, 
~\cite{Carsten}
 for providing the dynamical Wilson 
configurations.
 This work is
supported in part by the  EU Integrated Infrastructure Initiative
Hadron Physics (I3HP) under contract RII3-CT-2004-506078 and by the
U.S. Department of Energy (D.O.E.) 
Office of Nuclear Physics under contract DF-FC02-94ER40818.
This research used resources of the John von Neumann Institute of Computing
in Germany and of the National Energy Research Scientific
Computing Center supported by the Office of Science of the U.S.
D.O.E under Contract No. DE-AC03-76SF00098.


\begin{thebibliography}{99}
\bibitem{footnote}Although
$Lm_\pi=3.6$ in constrast to $Lm_\pi > 4.5$ in all other cases 
 the nucleon and $\Delta$ masses and form factors do not show
any qualitative change in behaviour as a function 
of $m_\pi$.
\end{thebibliography}
\end{document}